\documentclass[letterpaper]{article} 
\usepackage{aaai2026}  
\usepackage{times}  
\usepackage{helvet}  
\usepackage{courier}  
\usepackage[hyphens]{url}  
\usepackage{graphicx} 
\usepackage{natbib}  
\usepackage{caption} 
\nocopyright        
\usepackage{algorithm}
\usepackage{algorithmic}
\usepackage{amsmath}  
\usepackage{amssymb}  
\usepackage{booktabs} 

\title{Holmes: An Evidence-Grounded LLM Agent for Auditable DDoS Investigation in Cloud Networks}

\author {
    Haodong Chen\textsuperscript{\rm 1},
    Ziheng Zhang\textsuperscript{\rm 1},
    Jinghui Jiang\textsuperscript{\rm 1},
    Xuanhao Liu\textsuperscript{\rm 1},
    Qiang Su\textsuperscript{\rm 1,}\thanks{Corresponding author.},
    Qiao Xiang\textsuperscript{\rm 1,}\footnotemark[1]
}
\affiliations {
    \textsuperscript{\rm 1}Xiamen University, Xiamen, China
}

\newcommand{\sysname}{Holmes}

\begin{document}

\maketitle

\begin{abstract}
Cloud environments face frequent DDoS threats due to centralized resources and broad attack surfaces. Modern cloud-native DDoS attacks further evolve rapidly and often blend multi-vector strategies, creating an operational dilemma: defenders need wire-speed monitoring while also requiring explainable, auditable attribution for response. Existing rule-based and supervised-learning approaches typically output black-box scores or labels, provide limited evidence chains, and generalize poorly to unseen attack variants; meanwhile, high-quality labeled data is often difficult to obtain in cloud settings. We present \sysname (\textbf{DDoS Detective}), an \textbf{LLM-based DDoS detection agent} that reframes the model as a \emph{virtual SRE investigator} rather than an end-to-end classifier. \sysname couples a funnel-like hierarchical workflow (counters/sFlow for continuous sensing and triage; PCAP evidence collection triggered only on anomaly windows) with an \textbf{Evidence Pack} abstraction that converts binary packets into compact, reproducible, high-signal structured evidence. On top of this evidence interface, \sysname enforces a structure-first investigation protocol and strict JSON/quotation constraints to produce machine-consumable reports with auditable evidence anchors. We evaluate \sysname on CICDDoS2019 reflection/amplification attacks and script-triggered flooding scenarios. Results show that \sysname produces attribution decisions grounded in salient evidence anchors across diverse attack families, and when errors occur, its audit logs make the failure source easy to localize, demonstrating the practicality of an LLM agent for cost-controlled and traceable DDoS investigation in cloud operations.
\end{abstract}

\section{Introduction}
\label{sec:intro}
The rapid proliferation of cloud computing has established it as the backbone of the modern digital economy. However, this centralization of resources has inherently made cloud environments a prime target for Distributed Denial-of-Service (DDoS) attacks. Unlike traditional volumetric attacks, modern cloud-native DDoS attacks are characterized by their extreme diversity and dynamic evolution, often mixing multi-vector strategies (e.g., L7 application floods blended with L3/L4 amplification) to evade detection. Traditionally, Cloud Service Providers (CSPs) have relied heavily on expert-defined heuristic rules and manual intervention to mitigate these threats. While effective for known attack signatures, this reactive and labor-intensive approach struggles to keep pace with the increasing sophistication and polymorphism of modern attack vectors.

To automate defense mechanisms and reduce human dependency, recent research has pivoted towards data-driven approaches, utilizing Machine Learning (ML) and Deep Neural Networks (DNN) to classify malicious traffic. Notable works, such as TrafficLLM~\citep{cui2025trafficllm}, treat traffic analysis as a sequence modeling problem, training classifiers on massive datasets of historical, labeled network traffic. These methods aim to learn the statistical distributions of attack patterns, attempting to provide an automated binary or multi-class verdict for incoming traffic flows.

Despite their performance on benchmarks, these supervised learning paradigms face three critical hurdles in real-world cloud deployment.
First, they operate primarily as \textbf{black boxes}. A neural classifier typically outputs a probabilistic score (e.g., "99\% malicious") without offering an explainable root cause or the "evidence chain" required for security operations.
Second, they suffer from \textbf{poor generalization} on zero-day attacks. When encountering unseen traffic patterns, these models tend to blindly map inputs to the nearest learned distribution, leading to confident but erroneous predictions.
Third, and perhaps most uniquely to the cloud context, is the \textbf{scarcity of high-quality data}. Due to strict privacy regulations and commercial sensitivity, obtaining large-scale, annotated DDoS datasets for training supervised models is often impractical for CSPs.

We argue that the next generation of network security requires a paradigm shift from simple classification to \textit{evidence-driven investigation}. The emergence of Large Language Models (LLMs) offers a promising avenue, not as direct traffic classifiers, but as \textbf{Virtual Site Reliability Engineers (SREs)}. Unlike black-box models, an LLM-based agent can mimic the human workflow: observing symptoms, orchestrating diagnostic tools, and formulating hypotheses based on retrieved facts. However, relying solely on the probabilistic reasoning of LLMs is risky due to potential unfaithfulness. Therefore, we propose \textbf{\sysname}, an agent framework that constrains the LLM within a closed-loop workflow of \textit{tool execution} and \textit{evidence validation}, ensuring that every verdict is not just a probability score, but a verifiable conclusion backed by an evidence chain.

However, realizing such an effective Agent in production cloud environments presents several non-trivial technical challenges:

\begin{itemize}
    \item \textbf{Challenge 1: The Modality Gap between Binary Traffic and Semantic Reasoning.}
    Network traffic is inherently high-volume, numerical, and low-level (e.g., packet bytes, statistical distributions), whereas LLMs excel at processing semantic text. Feeding raw PCAP files directly into an LLM is computationally prohibitive and semantically opaque. How to abstract high-frequency traffic data into a semantic representation that serves as \textit{interpretable evidence} rather than just statistical features is a primary challenge.

    \item \textbf{Challenge 2: Unfaithful Reasoning and Tool Misinterpretation.}
    Security operations require rigorous accuracy. A general-purpose LLM may hallucinate non-existent parameters or misinterpret raw tool outputs (e.g., confusing benign TCP retransmissions with a flood). Designing a mechanism to strictly ground the Agent's reasoning in factual observations—preventing "security hallucinations"—is critical for operator trust.

    \item \textbf{Challenge 3: The Operational Efficiency Paradox.}
    There is a fundamental conflict between the wire-speed throughput required for DDoS monitoring and the high latency of LLM inference. Deploying a sophisticated reasoning agent on every traffic flow is economically and computationally infeasible. The challenge lies in designing a system that can continuously monitor massive traffic volumes while restricting expensive LLM investigations to only the necessary "moment of crisis."
\end{itemize}

To address these challenges, we propose \sysname, encompassing three core designs:

\textbf{First, to resolve the operational efficiency paradox (Challenge 3), we implement a Hierarchical Detection Workflow.}
Instead of an end-to-end LLM approach, our system employs a "funnel-like" pipeline. It utilizes lightweight interface counters and packet sampling (sFlow) as continuous sentinels, activating the deep reasoning Agent \textit{on-demand} only when specific anomaly thresholds are breached. This design effectively filters the vast majority of background traffic, allowing the system to focus expensive inference resources solely on complex, high-risk incidents within deployable latency budgets.

\textbf{Second, to bridge the modality gap (Challenge 1), we introduce a Semantic Evidence Abstraction mechanism.}
We move beyond simple feature engineering by constructing a structured \textbf{"Evidence Pack."} A specialized toolchain extracts "Primary Anchors"—such as protocol-specific headers, payload entropy, and characteristic strings—transforming unintelligible binary streams into a dense, textual representation. Crucially, this abstraction is designed not just for classification, but to provide \textit{quotable artifacts} that align with the LLM's reasoning process.

\textbf{Finally, to mitigate unfaithful reasoning (Challenge 2), we establish an Evidence-Grounded Reasoning paradigm.}
We enforce a strict "Quote Rule" via prompt engineering: the Agent is prohibited from open-ended generation and must explicitly cite verbatim substrings from the Evidence Pack to support its verdict. This constraint significantly reduces hallucinations and ensures that the Agent's output transforms from a probabilistic guess into a transparent, verifiable chain of evidence.

In summary, our main contributions are:
\begin{itemize}
    \item We propose a \textbf{Hierarchical Agent Framework} that reconciles the conflict between the slow inference of large models and the real-time requirements of cloud defense, making LLM-based security practically deployable.
    \item We design an \textbf{Anchor-Based Modality Alignment} method that converts binary packets into semantic Evidence Packs, enabling LLMs to comprehend network traffic without retaining massive raw data.
    \item We implement a \textbf{Quote-Constrained Reasoning} mechanism that enforces factual grounding, producing interpretable and verifiable incident reports suitable for professional security operations.
\end{itemize}

\section{Related Work}
\label{sec:related-work}

This paper lies at the intersection of cloud DDoS operations, traffic understanding, and LLM-based agentic systems. We review related work along the operational pipeline---from classical detection and mitigation, to ML/LLM-based traffic analysis, and finally to trustworthy, evidence-grounded LLM agents.

\begin{table*}[t] 
\centering
\small
\setlength{\tabcolsep}{12pt} 
\begin{tabular}{l|c|c|c|c}
\hline
\textbf{System/Line} & \textbf{Goal} & \textbf{LLM Role} & \textbf{Evidence} & \textbf{On-demand?} \\
\hline
Supervised ML/DL IDS & classify & none / classifier & score/label & often always-on \\
TrafficLLM-style~\citep{cui2025trafficllm} & classify/gen & traffic model & learned repr. & typically always-on \\
IDS-Agent-style~\citep{li2024idsagent} & detect+explain & agent & free-form exp. & task-dependent \\
ShieldGPT-style~\citep{wang2024shieldgpt} & mitigate & assistant/agent & prompts+repr. & task-dependent \\
\hline
\textbf{\sysname (ours)} & investigate & virtual SRE agent & \textbf{quote-grounded} & \textbf{yes} \\
\hline
\end{tabular}
\caption{Positioning of \sysname versus representative related lines of work.}
\label{tab:rw-compare}
\end{table*}

\subsection{Cloud DDoS Detection and Mitigation Pipelines}
Cloud service providers (CSPs) typically employ multi-layer defense pipelines, combining always-on lightweight telemetry (e.g., interface counters and packet sampling) with heavier on-demand analysis and mitigation actions. Such tiered workflows are operationally appealing because they decouple wire-speed monitoring from expensive deep inspection and attribution. However, in practice, incident triage and root-cause analysis still rely heavily on expert-driven heuristics and manual investigation, which can become brittle under rapidly evolving, multi-vector attacks.
Recent LLM-assisted mitigation systems (e.g., ShieldGPT~\citep{wang2024shieldgpt}) highlight the potential of LLMs for explanation and response guidance in DDoS operations.
\textbf{In contrast, \sysname focuses on automating the \emph{investigation} stage} of the pipeline: rather than replacing all mitigation controls, our goal is to mimic SRE-style diagnosis using tool-orchestrated reasoning while producing auditable evidence chains.

\subsection{Behavior- and Statistics-based DDoS Detection}
Prior to learning-based detectors, DDoS defense largely relied on \emph{behavioral signatures} and \emph{statistical deviations} from network/application telemetry.
On the behavior side, representative systems target distinct attack families—e.g., bursty volumetric flooding (ALBUS~\citep{scherrer2023albus}), application-layer flash-crowd confusion (FRADE~\citep{tandon2021flashcrowd}), exploit-style low-rate DoS via per-connection resource abuse (Leader~\citep{tandon2023leader}), and evasive temporal/pulse-wave dynamics (ACC-Turbo~\citep{alcoz2022accturbo}).
Behavioral reasoning is also central for link-flooding attacks (LFA), where defenses exploit routing/topology perturbations and attacker reaction patterns~\citep{liaskos2018topology,liaskos2016framework,rezapour2021rlshield,gkounis2016interplay,kang2016spiffy,ma2019patrolling}.

Statistics-based approaches instead detect \emph{distribution shifts} in compact metrics, with entropy (e.g., joint entropy) being a classic signal (JESS~\citep{kalkan2018jess}).
For LFAs, systems combine flow statistics, multi-vantage measurements, and correlation analysis for localization and attribution~\citep{xue2014linkscope,zheng2018realtime}, while newer switch-assisted designs coordinate at scale with network-wide views~\citep{xing2021ripple,zhou2023mew}.
To meet high-throughput and high-cardinality constraints, many designs adopt sketching as a core primitive, including Count-Min/Count Sketch and universal sketching frameworks~\citep{zhang2020poseidon,liu2021jaqen}, with adaptations to NDN-style persistence detection~\citep{xu2022lieffifm}.
Programmable data planes further enable in-switch metric computation (e.g., normalized entropy) and reuse of general monitoring sketches~\citep{ding2021p4nentropy,liu2016univmon}.

Despite long-standing success, two limitations remain for cloud-grade DDoS investigation: (i) threshold tuning and limited \emph{auditable, incident-oriented evidence}, and (ii) weak support for \emph{multi-stage} triage from coarse anomalies to packet-level root causes.
\textbf{\sysname addresses these gaps via three design choices}:
(i) a hierarchical on-demand workflow that activates deep investigation only upon anomalies,
(ii) a semantic evidence abstraction that converts low-level traffic into compact, quote-able Evidence Packs, and
(iii) an evidence-grounded reasoning protocol that enforces explicit citations to observed substrings.

\subsection{Large Language Models for Network Traffic Understanding}
Recent efforts explore adapting LLMs to the traffic domain by tokenizing traffic traces and learning generic traffic representations for downstream detection and generation tasks, such as TrafficLLM~\citep{cui2025trafficllm}. While these works significantly advance open-set robustness and cross-scenario generalization, they still largely cast the problem as \emph{traffic-to-label} (or \emph{traffic-to-text}) modeling, with limited emphasis on producing an auditable evidence chain tailored to operational incident response.
\textbf{Our perspective differs in role and objective}: \sysname does not treat the LLM as a direct traffic classifier; instead, it treats the LLM as a virtual SRE that orchestrates domain tools and synthesizes a verifiable, evidence-grounded decision.

\subsection{LLM-based Cybersecurity Agents}
LLM-powered agents have emerged as a promising paradigm for explainable security analysis, including agentic IDS frameworks such as IDS-Agent~\citep{li2024idsagent}, which iteratively reasons over observations, retrieves knowledge, and takes actions to produce predictions with explanations. Similarly, LLM-driven systems for DDoS mitigation (e.g., ShieldGPT~\citep{wang2024shieldgpt}) emphasize domain knowledge injection and role prompting to generate mitigation recommendations.
Despite encouraging progress, two gaps remain for cloud-grade DDoS investigation. First, many systems rely on free-form natural-language explanations that are difficult to audit against raw telemetry. Second, their inference cost can be difficult to justify under always-on, high-throughput production traffic.
\textbf{\sysname addresses these gaps via three design choices}:
(i) a hierarchical on-demand workflow that activates deep investigation only upon anomalies,
(ii) a semantic evidence abstraction that converts low-level traffic into compact, quote-able Evidence Packs, and
(iii) an evidence-grounded reasoning protocol that enforces explicit citations to observed substrings.

\section{Design}
\label{sec:design}

\begin{figure}
    \centering
    \includegraphics[width=\linewidth]{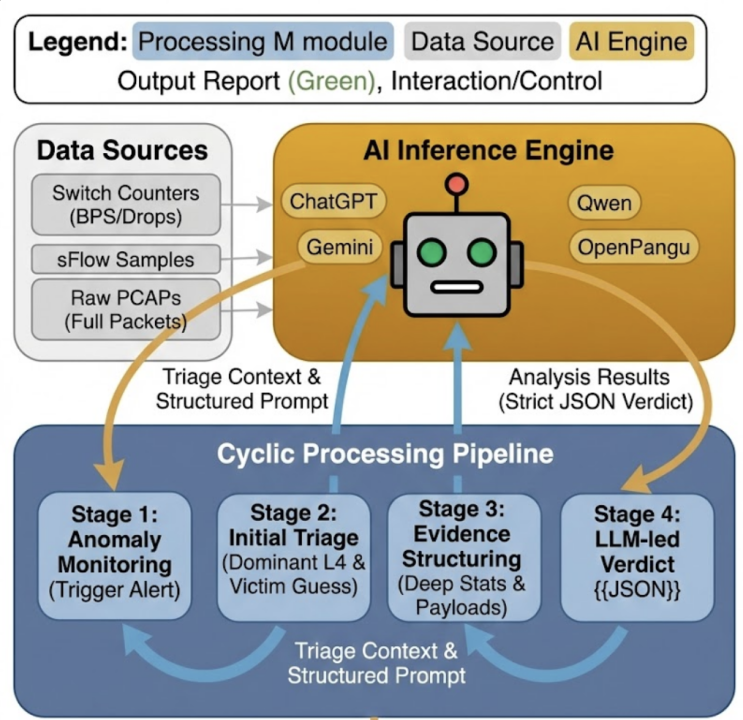}
    \caption{DDoS Detective Overview}
    \label{fig:overview}
\end{figure}

\sysname is built on a core principle: instead of treating an LLM as a black-box end-to-end traffic classifier, we reshape it into a \textbf{Virtual Site Reliability Engineer (Virtual SRE)} that conducts \emph{evidence-driven investigation}. Under the prerequisite of wire-speed monitoring in cloud environments, \sysname activates expensive reasoning only \emph{on demand}, transforms low-level binary packets into \emph{auditable semantic evidence}, and finally produces a \emph{verifiable} verdict and actionable response suggestions under strict evidence constraints. As illustrated in Figure~\ref{fig:overview}, the design consists of three core components, each directly addressing one of the agent challenges.

\subsection{Hierarchical Perception and On-Demand Investigation}
\label{sec:design-hierarchical}
To reconcile the conflict between always-on, wire-speed monitoring and the high latency/cost of LLM inference, we design a funnel-like hierarchical workflow that limits deep reasoning to critical moments. The workflow contains three abstract layers:

\textbf{L1: Continuous Telemetry.}
\sysname continuously monitors lightweight data-plane telemetry signals such as interface counters (e.g., bandwidth and queue drops). This layer acts as a low-cost ``gatekeeper'' to filter the vast majority of benign/background time windows (in our evaluation, most windows never trigger deep investigation), while preserving high recall for anomaly candidates.

\textbf{L2: Lightweight Triage.}
Once L1 triggers an anomaly threshold, \sysname performs lightweight triage using packet sampling (e.g., sFlow). The goal of L2 is not to make a final attack decision, but to \emph{route} the incident to the correct evidence-collection branch by quickly determining the \emph{dominant L4 protocol} (UDP/TCP/MIXED) and estimating the \emph{likely victim} (destination). This routing step is essential to avoid unnecessary or misdirected deep inspection.

\textbf{L3: On-Demand Investigation.}
Only after L2 completes the triage routing does \sysname activate the expensive agentic path. It performs budgeted packet-level evidence extraction from the PCAP within the current time window (via tshark) and invokes the LLM agent for investigation. To prevent repeated resource consumption on the same incident window, \sysname additionally introduces \emph{cooldown} and \emph{deduplication} mechanisms that suppress redundant investigations under bursty alert streams.

Overall, this hierarchical workflow decouples monitoring breadth from reasoning depth, making LLM-based incident investigation economically feasible in production-like cloud settings.

\subsection{Semantic Evidence Abstraction and Evidence Pack}
\label{sec:design-evidence}
Raw PCAPs are large, binary, and semantically opaque, which makes direct LLM consumption both expensive and unreliable. To bridge this modality gap, \sysname introduces a \textbf{Semantic Evidence Abstraction} module that converts low-level packet traces into a structured, LLM-facing interface called the \textbf{Evidence Pack}. The Evidence Pack is designed for \emph{high signal density}, \emph{quote-ability}, \emph{reproducibility}, and \emph{budgeted generation}.

\textbf{Representative Sampling.}
Instead of full PCAP scanning, \sysname extracts a small number of representative packets that preserve the \emph{structural fingerprints} of the incident. In UDP-dominant cases, \sysname prioritizes the primary traffic cluster (PRIMARY cluster) and samples a few representative payloads based on the dominant \texttt{udp.length} modes. In TCP-dominant cases, the abstraction focuses on TCP flag distributions as the most direct evidence for SYN/ACK floods, optionally complemented by a small number of packet samples when needed.

\textbf{Structural Fingerprinting (Ports are not identity).}
The abstraction emphasizes payload structure cues and readable artifacts rather than unreliable port-based assumptions (ports may be randomized). For each sampled packet, \sysname derives compact indicators such as printable-character ratio, Shannon entropy, and short ASCII excerpts/anchors. These artifacts enable the agent to reason over structure styles (e.g., \texttt{http\_like}, \texttt{asn1\_oid\_like}, or semi-structured key/value lists) instead of overfitting to superficial metadata.

\textbf{Budget-Aware Generation.}
To ensure deterministic cost and latency, \sysname applies strict caps on evidence extraction, including upper bounds on tshark scan packets, hexdump lines/characters, and the number of samples. This makes Evidence Pack size and generation time predictable, preventing evidence collection itself from becoming a bottleneck.

By exposing compact, structured, and quote-able artifacts, the Evidence Pack serves as an auditable ``evidence API'' between high-throughput traffic data and LLM reasoning.

\subsection{Evidence-Grounded Reasoning with Quote and Format Constraints}
\label{sec:design-grounded}

LLM agents may produce ungrounded claims or misread tool outputs in safety-critical security operations. \sysname therefore enforces an \textbf{Evidence-Grounded Reasoning} paradigm that restricts the agent to an \emph{observe--verify--conclude} loop, where conclusions must be supported by tool-derived evidence and emitted in a machine-consumable form.

\paragraph{Structure-first multi-step investigation protocol.}
We define a structure-first investigation protocol (Route-1) that decomposes decision making into three stages:
(i) infer the \emph{payload structure style} from PRIMARY evidence (e.g., whether the payload is noise-like, HTTP-like, or ASN.1/OID-like),
(ii) determine the \emph{attack family} (reflection/amplification vs.\ direct flood), and
(iii) decide the \emph{attack type} under strict gates (e.g., TCP-dominant incidents must not be labeled as reflection; UDP-dominant incidents must not be labeled as SYN/ACK floods).
This staged protocol prevents the agent from jumping to conclusions based on weak or spurious cues and makes the decision process auditable.

\paragraph{Prompt Contract: grounding and output specification.}
To make the above protocol reproducible and robust, we encode the agent behavior as a concise \textbf{Prompt Contract} consisting of (1) hard invariants, (2) the Route-1 protocol, and (3) a strict JSON schema.
Figure~\ref{fig:prompt-contract} shows a simplified (paper-version) prompt contract; the full prompt with extended guidance and examples is included in the supplementary material.

\begin{figure}[t]
\centering
\fbox{
\begin{minipage}{0.96\columnwidth}
\footnotesize
\textbf{Prompt Contract (Simplified).} Note: output \emph{strict JSON only}.

\vspace{0.35em}
\textbf{Invariants.}
\begin{itemize}
  \item \textbf{Single-source:} use only the Evidence Pack in this message (no external facts).
  \item \textbf{Ports are not identity:} do not use port numbers to identify protocols (ports may be randomized).
  \item \textbf{Quote Rule:} each item in \texttt{key\_evidence} must include $\ge$1 \emph{verbatim} substring from the Evidence Pack wrapped in backticks.
\end{itemize}

\vspace{0.35em}
\textbf{Route-1 (structure-first).}
\begin{enumerate}
  \item Decide \texttt{payload\_style} from PRIMARY samples:
  \texttt{random\_noise} $|$ \texttt{http\_like} $|$ \texttt{asn1\_oid\_like} $|$
  \texttt{kv\_semicolon\_list} $|$ \texttt{text\_banner\_like} $|$ \texttt{mixed\_or\_unclear}.
  \item Infer \texttt{attack\_family}:
  \texttt{Reflection/Amplification} $|$ \texttt{Direct Flood} $|$ \texttt{Mixed} $|$ \texttt{Unknown}.
  \item Decide \texttt{attack\_type} with gates:
  \begin{itemize}
    \item If \texttt{dominant\_l4\_used=TCP}: choose only \texttt{SYN Flood}, \texttt{ACK Flood}, \texttt{HTTP/2 Rapid Reset}, or \texttt{Unknown}.
    \item If \texttt{dominant\_l4\_used=UDP}: do not output \texttt{SYN Flood}/\texttt{ACK Flood}; use reflection labels only if \emph{quotably} supported by payload structure.
    \item If structure is unclear, prefer \texttt{Unknown} rather than forcing a protocol.
  \end{itemize}
\end{enumerate}

\vspace{0.35em}
\textbf{Strict JSON schema (no extra keys).}
\texttt{verdict}, \texttt{attack\_family}, \texttt{attack\_type},
\texttt{analysis\_trace} (incl.\ \texttt{payload\_style}, \texttt{primary\_samples\_checked}, \texttt{decision}),
\texttt{key\_evidence} (2--6 quoted strings),
\texttt{reasoning} (1--3 sentences incl.\ a quote),
\texttt{recommended\_actions} (3--6 items),
\texttt{confidence} $\in [0,1]$.
\end{minipage}
}
\caption{Simplified prompt contract used by \sysname. It formalizes evidence grounding (Quote Rule), a structure-first investigation protocol (Route-1), and strict JSON outputs for downstream consumption.}
\label{fig:prompt-contract}
\end{figure}

\paragraph{Format constraints and self-correction.}
To make agent outputs reliably consumable by downstream defense workflows, \sysname requires strict JSON-only responses under the schema above. If the output fails parsing or field validation (e.g., missing fields or invalid types), \sysname triggers a feedback-based self-correction loop that provides validation errors and requests a corrected JSON response. This improves robustness under production variability without requiring changes to the evidence generator.

\paragraph{Audit logging.}
Finally, \sysname logs the anomaly context, triage results, the full Evidence Pack, and both intermediate and final agent outputs for incident review and continuous improvement. Together, these constraints transform the LLM from a free-form generator into a constrained investigator that produces verifiable, auditable incident reports under practical deployment budgets.

\section{Experiments}
\label{sec:experiments}

We evaluate \sysname (\textbf{DDoS Detective}) on a replay-driven simulation platform that replays (i) reflection/amplification attacks from CICDDoS2019 and (ii) script-triggered synthetic flooding attacks. During replay, we continuously collect switch interface counters, sFlow samples, and time-sliced PCAPs; \sysname runs online and triggers an \emph{incident window} whenever the counter monitor detects an anomaly. Each incident is then investigated via the LangGraph workflow (monitor $\rightarrow$ triage $\rightarrow$ evidence $\rightarrow$ detective), where the detective agent is powered by an OpenPangu-7B LLM \cite{chen2025pangu}.
20

\begin{table*}[t]
\centering
\scriptsize
\setlength{\tabcolsep}{3.2pt}
\renewcommand{\arraystretch}{1.15}
\begin{tabular}{l|l|c|c|p{0.36\textwidth}|c|c|c|c}
\hline
\textbf{Inc. (T)} & \textbf{PCAP slice} & \textbf{Victim} & \textbf{Dom. L4 (score)} &
\textbf{Primary evidence anchor(s) from logs} &
\textbf{GT} & \textbf{Pred} & \textbf{OK} & \textbf{Conf.} \\
\hline
29s  & 27s--44s   & 10.0.0.8 & UDP (0.981) &
noise-like payload; printable\_ratio$\approx$0.353 \newline
entropy$\approx$7.53 &
UDP-F & UDP-F & Y & 0.90 \\
\hline
213s & 210s--233s & 10.0.0.8 & TCP (0.971) &
tshark flags: ack\_only\_ratio$\approx$0.977 \newline
ack\_ratio$\approx$0.977 &
ACK-F & SYN-F & N & 0.95 \\
\hline
260s & 258s--277s & 10.0.0.8 & TCP (0.992) &
tshark flags: syn\_only\_ratio$\approx$0.995 \newline
\texttt{HTTP/1.1 301 Moved Permanently} (TCP payload) &
SYN-F & SYN-F & Y & 0.75 \\
\hline
326s & 322s--349s & 10.0.0.7 & UDP (0.979) &
DNS-like domain in payload: \texttt{aids.gov} \newline
(multiple udp.length modes observed) &
DNS-R & DNS-R & Y & 0.90 \\
\hline
486s & 485s--554s & 10.0.0.7 & UDP (0.881) &
NetBIOS anchors: \texttt{\_\_MSBROWSE\_\_}, \texttt{WORKGROUP} \newline
(top\_src\_share$\approx$0.97) &
NetBIOS-R & UDP-F & N & 0.85 \\
\hline
577s & 577s--609s & 10.0.0.7 & UDP (0.952) &
LDAP/CLDAP OID: \texttt{1.2.840.113556} \newline
attrs: \texttt{supportedLDAPVersion}, \texttt{isGlobalCatalogReady1} &
LDAP-R & LDAP-R & Y & 0.95 \\
\hline
633s & 633s--657s & 10.0.0.7 & UDP (0.958) &
SNMP anchor: \texttt{public} \newline
banner: \texttt{View-based Access Control Model for SNMP} &
SNMP-R & SNMP-R & Y & 0.80 \\
\hline
691s & 691s--732s & 10.0.0.7 & UDP (0.924) &
MSSQL KV fields: \texttt{ServerName;}, \texttt{InstanceName;} \newline
\texttt{MSSQLSERVER} appears in PRIMARY samples &
MSSQL-R & Other-R & N & 0.93 \\
\hline
745s & 744s--760s & 10.0.0.7 & UDP (0.984) &
SSDP/UPnP headers: \texttt{HTTP/1.1 200 OK}, \texttt{LOCATION:} \newline
\texttt{gatedesc.xml}, \texttt{UPnP} (HTTP-like over UDP) &
SSDP-R & Other-R & N & 0.95 \\
\hline
793s & 792s--807s & 10.0.0.7 & UDP (0.975) &
MSSQL KV anchors: \texttt{ServerName;}, \texttt{InstanceName;} \newline
\texttt{IsClustered;} appears in PRIMARY samples &
MSSQL-R & MSSQL-R & Y & 0.92 \\
\hline
\end{tabular}
\caption{\textbf{Two-column unified incident table} for representative windows from one 15-minute replay run.
\textbf{GT} is the replay scenario label (CICDDoS2019 reflection/amplification families or script-triggered floods), and each GT entry is verifiable from the logged Evidence Pack anchors shown above.
\textbf{Pred} is the final attack-type attribution in the generated incident report.
Abbreviations: UDP-F=UDP Flood, SYN-F=SYN Flood, ACK-F=ACK Flood, DNS-R=DNS Reflection, NB-R=NetBIOS Reflection, LDAP-R=LDAP/CLDAP Reflection, SNMP-R=SNMP Reflection, MSSQL-R=MSSQL Reflection, SSDP-R=SSDP/UPnP Reflection, Other-R=generic reflection label in the final report.}
\label{tab:incident-unified-2col}
\end{table*}

\subsection{Experimental Setup}
\label{sec:exp-setup}

\textbf{System implementation.}
DDoS Detective is implemented as a LangGraph state machine that orchestrates four stages: (1) counter-based anomaly monitoring, (2) sFlow triage, (3) budgeted PCAP evidence abstraction via tshark, and (4) an LLM-led incident investigation that outputs a strict JSON verdict and an auditable report.

\textbf{LLM backend.}
We deploy the OpenPangu 7B model as the detective LLM and invoke it through an OpenAI-compatible API endpoint (temperature set to 0 for deterministic behavior). The model is constrained by (i) a structure-first investigation protocol and (ii) strict JSON output requirements described in \S\ref{sec:design-grounded}.

\textbf{Traffic sources and replay.}
We use reflection/amplification attack traffic from CICDDoS2019 (including DNS / NetBIOS / SNMP / LDAP(CLdap) / MSSQL / SSDP(UPnP) / NTP families) and generate synthetic flooding attacks (UDP Flood / SYN Flood / ACK Flood) using scripts. All traffic is replayed in a simulation environment; counters, sFlow, and PCAP slices are collected during replay to form the inputs to DDoS Detective. One full run lasts \textbf{15 minutes} and yields multiple counter-triggered incident windows.

\textbf{Evaluation protocol.}
The unit of evaluation is an \emph{incident window} determined by the counter anomaly trigger and mapped to a time-sliced PCAP. For each incident, we report: (i) sFlow triage summary (dominant L4 protocol and victim guess), (ii) structured Evidence Pack produced by tshark-based abstraction, and (iii) the LLM verdict with confidence and evidence highlights. Our focus is on end-to-end operational investigation outputs (attack type attribution + evidence traceability), rather than packet-level classification.

\subsection{Results}
\label{sec:exp-results}

Table~\ref{tab:incident-unified-2col} summarizes representative incident windows from one 15-minute replay run, including triage signals, PRIMARY evidence anchors from the Evidence Pack, and the final attribution outcomes.

\subsubsection{End-to-end incident outcomes}
We summarize each incident window in a single two-column table that jointly reports:
(i) triage signals (dominant L4 and victim),
(ii) salient PRIMARY evidence anchors extracted into the Evidence Pack, and
(iii) the final attribution outcome (ground-truth label from replay, predicted type, and whether the attribution matches the replay label).
Table~\ref{tab:incident-unified-2col} presents this unified view, enabling direct auditing from evidence anchors to the agent’s decision.

\noindent\textbf{Takeaway.}
Across these incidents, DDoS Detective consistently attributes reflection/amplification attacks when PRIMARY samples expose strong protocol-structural anchors (e.g., LDAP OIDs/attributes, SNMP community/banner strings, MSSQL semicolon-delimited fields, and SSDP/UPnP HTTP-like headers over UDP), while noise-like payloads are attributed as direct flooding.
Because Table~\ref{tab:incident-unified-2col} records \emph{triage signals}, \emph{evidence anchors}, and \emph{final decisions} in one place, disagreements between ground truth and prediction can be audited directly by inspecting the logged anchors.

\subsubsection{Evidence traceability (auditable attribution)}
A key advantage of DDoS Detective is that each verdict is accompanied by an Evidence Pack that exposes \emph{human-auditable anchors} extracted from PRIMARY samples. Below we show concise evidence-to-verdict trace snippets (Evidence Pack $\rightarrow$ output) that enable rapid operator verification:

\begin{quote}\small
\textbf{LDAP/CLDAP reflection (evidence anchors).}
Evidence Pack contains OIDs and directory attributes such as \texttt{1.2.840.113556} and \texttt{supportedLDAPVersion} / \texttt{isGlobalCatalogReady1}, and the agent outputs \emph{LDAP/CLDAP Reflection} with these anchors as key evidence.

\textbf{SNMP reflection (evidence anchors).}
Evidence Pack contains \texttt{public} and banner-like text \texttt{View-based Access Control Model for SNMP}, enabling an \emph{SNMP Reflection} attribution.

\textbf{MSSQL reflection (evidence anchors).}
Evidence Pack shows semicolon-delimited fields such as \texttt{ServerName; ... InstanceName; ... MSSQLSERVER}, supporting an \emph{MSSQL Reflection} attribution.

\textbf{SSDP/UPnP reflection (evidence anchors).}
Evidence Pack shows HTTP-like UDP payloads including \texttt{HTTP/1.1 200 OK} with \texttt{LOCATION:} and \texttt{gatedesc.xml}, matching a classic UPnP/SSDP reflection structure.
\end{quote}

These anchors make the investigation \emph{traceable by design}: analysts can directly locate the quoted strings in the Evidence Pack and validate whether the attribution follows from observed payload structures, rather than trusting a black-box score.

\subsection{Case Studies and Failure Analysis}
\label{sec:exp-cases}

\subsubsection{Case Study: LDAP/CLDAP reflection (high-confidence and strongly grounded)}
We highlight the incident at \textbf{T=577.0s} (PCAP slice \texttt{577s--609s.pcap}) as a representative success case that demonstrates the end-to-end investigation loop.

\textbf{(1) Triage.}
The sFlow triage identifies a UDP-dominant incident (dominance score $\approx$ 0.952) and predicts the victim as \texttt{10.0.0.7}. This immediately routes the investigation to the UDP evidence branch.

\textbf{(2) Evidence Pack: PRIMARY samples expose protocol structure.}
DDoS Detective extracts representative UDP samples (PRIMARY source and dominant \texttt{udp.length} modes). The PRIMARY samples contain high-signal, protocol-specific structure cues: the payload includes multiple OID-like strings and directory attributes such as
\texttt{1.2.840.113556} and \texttt{supportedLDAPVersion}, as well as status-like fields including \texttt{isGlobalCatalogReady1} and \texttt{domainFunctionality1}. These artifacts are visible in the ASCII excerpts and corroborated by the hexdump.

\textbf{(3) Investigation output and auditable attribution.}
Given the ASN.1/OID-like structure and LDAP attribute anchors, the agent attributes the incident as \emph{LDAP/CLDAP Reflection} with high confidence. Importantly, the incident report surfaces the above substrings as key evidence, allowing an operator to verify the attribution by directly cross-checking the Evidence Pack.

\textbf{Why this works.}
This case illustrates the core design of DDoS Detective: (i) the funnel workflow concentrates analysis on anomaly windows; (ii) semantic evidence abstraction exposes compact, high-signal PRIMARY samples; and (iii) the LLM reasoning is guided by structure-first protocol cues (OID/attribute patterns), which generalize beyond port-based heuristics and remain auditable.

\subsubsection{Failure Analysis (brief): TCP incident with HTTP semantic interference}
We briefly discuss the incident at \textbf{T=260.0s} (PCAP slice \texttt{258s--277s.pcap}) where TCP is dominant and the Evidence Pack simultaneously contains (i) strong transport-layer signals (SYN-heavy flags) and (ii) application-layer HTTP-like payload text (e.g., \texttt{HTTP/1.1 301 Moved Permanently} and a \texttt{Location:} URL). In this mixed-evidence setting, the LLM output shows signs of semantic distraction (over-weighting HTTP-like strings while reasoning about a transport-layer flood type), which reflects a limitation of LLM robustness under competing cues. Crucially, DDoS Detective’s auditable logging makes the root cause easy to diagnose: the Evidence Pack exposes both the TCP flag statistics and the HTTP-like payload excerpt side-by-side with the model’s final decision, enabling rapid identification that the inconsistency stems from LLM interpretation rather than missing observability in the framework.

\section{Conclusion}
\label{sec:conclusion}

This paper presented \sysname (\textbf{DDoS Detective}), an LLM-based DDoS detection agent that treats the model as a \emph{virtual SRE investigator} rather than an end-to-end traffic classifier. DDoS Detective integrates (i) a funnel-like hierarchical workflow that keeps monitoring wire-speed while triggering deep investigation only on anomaly windows, (ii) an \textbf{Evidence Pack} abstraction that bridges binary traffic and LLM reasoning with compact, reproducible, high-signal evidence anchors, and (iii) evidence-grounded reasoning constraints (structure-first protocol with strict JSON/quotation rules) that make incident reports auditable and operationally actionable. 

Experiments on CICDDoS2019 reflection/amplification attacks and script-triggered flooding scenarios show that DDoS Detective can attribute diverse DDoS incidents with traceable evidence anchors, and its audit logs make misattributions easy to diagnose, exposing limitations of LLM reasoning rather than gaps in observability.



\bibliography{aaai2026}

\end{document}